# Wrapper/TAM Co-Optimization and constrained Test Scheduling for SOCs Using Rectangle Bin Packing

Hafiz Md. Hasan Babu, Md. Rafiqul Islam, Muhammad Rezaul Karim, Abdullah Al Mahmud and Md. Saiful Islam

Department of Computer Science & Engineering

University of Dhaka

Dhaka-1000, Bangladesh

Hafizbabu@hotmail.com,{rafik3203,r_karimcs,aamrubel,sohel_csdu}@yahoo.com

**Abstract**

This paper describes an integrated framework for SOC test automation. This framework is based on a new approach for Wrapper/TAM co-optimization based on rectangle packing considering the diagonal length of the rectangles to emphasize on both TAM widths required by a core and its corresponding testing time .In this paper, an efficient algorithm has been proposed to construct wrappers that reduce testing time for cores. Rectangle packing has been used to develop an integrated scheduling algorithm that incorporates power constraints in the test schedule. The test power consumption is important to consider since exceeding the system's power limit might damage the system.

**INTRODUCTION**

The development of microelectronic technology has lead to the implementation of system-on-chip (SOC), where a complete system, consisting of several application specific integrated circuit (ASIC), microprocessors, memories and other intellectual properties (IP) blocks, is implemented on a single chip. The increasing complexity of SOC has created many testing problems. The general problem of SOC test integration includes the design of TAM architectures, optimization of the core wrappers, and test scheduling. Test wrappers form the interface between cores and test access mechanisms (TAMs), while TAMs transport test data between SOC pins and test wrappers [3]. We address the problem of designing test wrappers and TAMs to minimize SOC testing time. While optimized wrappers reduce test application times for the individual cores, optimized TAMs lead to more efficient test data transport on-chip. Since wrappers influence TAM design, and vice versa, a co-optimization strategy is needed to jointly optimize the wrappers and the TAM for an SOC.

In this paper, we propose a new approach to integrated wrapper/TAM co-optimization and test scheduling based on a general version of rectangle packing considering diagonal length of the rectangles to be packed. The main advantage of the proposed approach is that it minimizes the test application time while considering test power limitation.

**RELATED WORK**

Most prior research has either studied wrapper design and TAM optimization as independent problems, or not addressed the issue of sizing TAMs to minimize SOC testing time [1,3,13] .Alternative approaches that combine TAM design with test scheduling [5,15] do not address the problem of wrapper design and its relationship to TAM optimization.

The first integrated method for Wrapper/TAM co-optimization was proposed in [10,11,12]. [10,12] are based on fixed-width TAMs which are inflexible and result in inefficient usage of TAM wires. An approach to wrapper/TAM co-optimization based on a generalized version of rectangle packing was proposed in [11]. This approach provides more flexible partitioning of the total TAM width among the cores. In a paper [16], a method is proposed to address the test power consumption, where the test time for a system with wrapped cores is minimized while test power limitations are considered and tests are assigned to TAM wires. [6] address the SOC test scheduling problem by proposing a test scheduling technique that minimizes the test application time while considering test power consumption and test conflicts.

**PROPOSED WRAPPER DESIGN**

The purpose of our wrapper design algorithm (Fig. 1) is to construct a set of wrapper chains at each core. A wrapper chain includes a set of the scanned elements (scan-chains, wrapper input cells and wrapper output cells). The test time at a core is given by:

$$T_{core} = \mathbf{p} \times [1+\max\{si,so\}] + \min\{si,so\}$$

where **p** is the number of test vectors to apply to the core and si (so) denotes the number of scan cycles required to load (unload) a test vector (test response)[10]. So, to reduce test time, we should minimize the longest wrapper chain (internal or external or both), *i.e.* max{si, so}**.** Recent research on wrapper design has stressed the need for balanced wrapper scan chains [3,10] to minimize the longest wrapper chain. Balanced wrapper scan chains are those that are as equal in length to each other as possible.

The proposed Wrapper_Design algorithm tries to minimize core testing time as well as the TAM width required for the test wrapper. The objectives are achieved by balancing the lengths of the wrapper scan chains and imposing an upper bound on the total number of scanned elements.

Our heuristic can be divided in two main parts; the first one for combinational cores and the second one for sequential cores. For combinational cores, there are two possibilities. If *I+O* (where *I* is the number of functional inputs and *O* the number of functional outputs) is below or equal to the TAM bandwidth limit, $W_{max}$, then nothing is done and the number of connections to the TAM is *I+O*. If *I+O is above* $W_{max}$, then some of the cells on the I/Os are chained together in order to reduce the number of needed connections to the TAM.

For sequential cores, at first an upper bound is specified (Upper_Bound). The internal scan chains are then sorted in descending order. After that, each internal scan chain is successively assigned to the wrapper scan chain, whose length after this assignment is closest to, but not exceeding the length of the upper bound. In our algorithm, a new wrapper scan chain is created only when it is not possible to fit an internal scan chain into one of the existing wrapper scan chains without exceeding the length of the upper bound. At last, functional inputs and outputs are added to balance the wrapper scan chains. Results of wrapper design algorithm are given in Table 1.

```
procedure Wrapper_Design (int W_max, Core C)
    {   //W_max =TAM width   ; //#SC=Total scan chain in Core C
    Total_Scan_Element= total  I/O +Σ  C.Scan_Chain_Length[i](1≤i<≤#SC);
    1. If  C.#SC=0           //combinational core
            If ( Total_Scan_Element <= W_max )
                    Assign one bit on every I/O wrapper cell;
            Else
                    Design W_max wrapper scan chains;
    2. Else                   //sequential core
        Mid_Lines = W_max / 2;
        Upper_Bound = Total_Scan_Element /Mid_Lines ;
        Sort the internal scan chains in descending order of their length;
        For each scan chain SC
           For each wrapper scan chain W already created
                If ( Length(W)+Length(SC)<=Upper_Bound )
                        Assign the scan chain to this wrapper scanchain W ;
            Else
                        Create a new Wrapper scan chain W_new ;
                        Assign the scan chain to this wrapper scanchain W_new ;
        Add functional I/O to balance the wrapper chains;
    }
```

**Fig. 1 Algorithm for wrapper design**

| TAM  size | TAM utilized ($TAM_u$) | Longest Scan chain |
|---|---|---|
| 50-64 | 47 | 521 |
| 48-49 | 39 | 1021 |
| 32-47 | 24 | 1042 |
| 24-31 | 16 | 1563 |
| 20-23 | 12 | 2084 |
| 16-19 | 10 | 2605 |
| 14-15 | 8 | 3126 |
| 12-13 | 7 | 3647 |
| 10-11 | 6 | 4689 |
| 8-9 | 5 | 5729 |
| 6-7 | 4 | 7809 |
| 4-5 | 3 | 11969 |
| 2-3 | 2 | 23789 |
| 1 | 1 | 24278 |

**Table 1. Result of Wrapper_Design for core 6 of p93791** [4]

## TAM DESIGN AND TEST SCHEDULING

The general integrated wrapper/TAM co-optimization and test scheduling problem that we address in this paper is as follows. We are given the total SOC TAM width and the test set parameters for each core. The set of parameters for each core includes the number of primary I/Os, test patterns, scan chains and scan chain lengths. The goal is to determine the TAM width and a wrapper design for each core, and a test schedule that minimizes the testing time for the SOC such that the following constraints are satisfied:

1. The total number of TAM wires utilized at any moment does not exceed $W_{max}$;
2. The maximum power dissipation value is not exceeded.

We formulate this problem as a progression of two problems of increasing complexity. These two problems are as follows:

*Problem 1:* wrapper/TAM co-optimization and test Scheduling

*Problem 2:* wrapper/TAM co-optimization and test    scheduling with power constraints.

In this section, we address *Problem 1* and show how wrapper/TAM co-optimization can be integrated with test scheduling. In the next section, we show how this problem is generalized to include power constraints- *Problem 2*.

*Problem 1:* determine the TAM width to be assigned and design a wrapper for each core and schedule the tests for the SOC in such a way that minimizes the total testing time as well as TAM width utilization and the total number of TAM wires utilized at any moment does not exceed total TAM width when a set of parameters for each core is given..

The concept of using rectangles for core test representation has been used before in [8,11,15]. Consider a SOC having N cores and let $R_i$ be the set of rectangles for core $i$, $1 \leq i \leq N$. Generalized version of rectangle packing problem *Problem-$_{RP}$1* is as follows: select a rectangle $R$ from $R_i$ for each set $R_i$, $1 \leq i \leq N$ and pack the selected rectangles in a bin of fixed height and unbounded width such that no two rectangle overlap and the width to which the bin is filled is minimized. Each rectangle selected is not allowed to be split vertically in our rectangle packing.

*Problem-$_{RP}$1* can be shown to be $N\rho$-hard. A special case of *Problem-$_{RP}$1*, in which the cardinality of each set $R_i$, $1 \leq i \leq N$ equals one, and no rectangles are allowed to be split, directly corresponds to the rectangle packing problem in [17]. Since the rectangle packing problem was shown to be $N\rho$-hard in [17] (by restriction to Bin Packing), *Problem-$_{RP}$1* is also $N\rho$-hard.

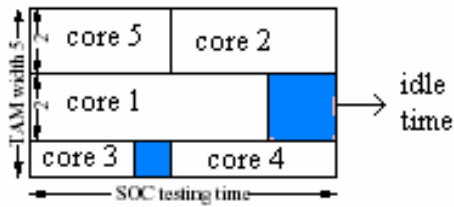

**Fig. 2 Example test schedule using rectangle packing**

We solve the *Problem 1* by generalized version of rectangle packing or two-dimensional packing *Problem-$_{RP}$1* .We use the Wrapper_Design algorithm to obtain the different test times for each core for varying values of TAM width. A set of rectangles for a core can now be constructed, such that the height of each rectangle corresponds to a different TAM width and the width of the rectangle represents the core test application time for this value of TAM width.

*Problem-$_{RP}$1* relates to *problem 1* as follows; see Fig. 2.The height of the rectangle selected for a core corresponds to the TAM width assigned to the core, while the rectangle width corresponds to its testing time. The height of the bin corresponds to the total SOC TAM width, and the width to which the bin is ultimately filled corresponds to the system testing time that is to be minimized. The unfilled area of the bin corresponds to the idle time on TAM wires during test. Furthermore, the distance between the left edge of each rectangle and the left edge of the bin corresponds to the begin time of each core test.

Our approach emphasizes on both testing time of a core and the TAM width required achieving that testing time by considering the diagonal length of rectangles. Diagonal length emphasizes on both testing time and TAM width since $DL=\sqrt{W^2 + H^2}$ where W, H, DL denotes width, height and diagonal length of the rectangles respectively. Consider three rectangles R[1] = {H=32, W=7.1, DL=32.78}, R[2] = {H=16, W=13.8, DL=21.13}, R[3] = {H=32, W=5.4, DL=32.45). Here if we take into account testing time(*W*), then we should pack R[2] first, followed by R[1] and R[3]. But when we consider diagonal lengths, we pack R[1], R[3], R[2] in sequence, and get the result that is extremely efficient.

Our approach minimizes TAM width utilization also by assigning $TAM_u$ wires to a core to achieve a specific testing time. For example, in our proposed Wrapper_Design, all TAM widths from 50 up to 64 result in the same testing time of 114317 cycles and same TAM width utilization($TAM_u$) of 47 for core 6 in p93791(Table 1). So, to achieve testing time of 114317 cycles $TAM_u$ value 47 is used in our proposed approach.

**POWER CONSTRAINED TEST SCHEULING**

This section, describes *Problem 2(*Integrated TAM design and power constrained test scheduling) in details and then formulates problem *Problem-$_{RP}$2*, a generalized version of Problem-$_{RP}$1 that is equivalent to *Problem 2*.

*Problem 2*: solve *Problem 1*, such that:

    1. The maximum power dissipation value $P_{max}$ is not exceeded.

Power constraints must be incorporated in the schedule to ensure that the power rating of the SOC is not exceeded during test.

*Problem 2* can be expressed in terms of rectangle packing as follows: consider a SOC having *N* cores, and:

    1. Let $R_i$ be the set of rectangles for core *i*, $1 \leq i \leq N$

    2. Let tests for core i have a power dissipation of $P_i$.

*Problem-$_{RP}$2:* solve *Problem-$_{RP}$1* ensuring that at any moment of time the sum of the $P_i$ values for the rectangles selected must not exceeded the maximum specified value $P_{max}$.

---

Algorithm Test_Scheduling ($W_{max}$, Core C[1...NC])

{

1. For each core C[i], construct a set of rectangles taking $TAM_u$ as rectangle height and its corresponding testing time as rectangle width such that $TAM_u \leq W_{max}$.

2. Find the smallest ($T_{min}$) among the testing time corresponding to MAX_$TAM_u$ of all cores.

3. For each core C[*i*], divide the width T[*i*] of all rectangles constructed in line 1 with $T_{min}$.

4. For each core C[*i*], calculate Diagonal Length  DL[*i*] = $\sqrt{(W[i])^2 + (T[i]^2)}$ where W[*i*] denotes MAX_TAM$_u$ and T[*i*] denotes corresponding reduced testing time.

5. Sort the Cores in descending order of their diagonal length calculated in line 4 and keep in list INITIAL[*NC*].

6. Next_Schedule_Time = current_Time = 0;

   Wavail = W$_{max}$;    // TAM available ;    Idle_Flag=False;

   //  peak_tam[c] is equal to MAX_TAM$_u$ of core c ;   //    PENDING is a queue.

7. While (INITIAL and PENDING not Empty)
   {
8. If (Wavail > 0 and Idle_Flag=False)
   {
   9. If (INITIAL is not empty)
   {             c=delete(INITIAL);
                 If  ( Wavail>=peak_tam[*c*] && no_powerConflict)
                     Update(c,peak_tam(*c*));
                 Else If(Possible_TAM >=0.5*peak_tam[c] && no_powerConflict)
                     Update(c, Possible_TAM);
                 Else
                     add(PENDING,c);
                 if(peak_tam[PENDING[*front*]] ≤ Wavail  && no_powerConflict)
                     Update(PENDING[*front*], peak_tam[PENDING[*front*]]);
                   delete(PENDING) ;
   }
   10. Else  //if INITIAL is empty
   {             If(peak_tam[PENDING[*front*]] ≤ Wavail && no_powerConflict)
                     Update(PENDING[*front*], peak_tam[PENDING[*front*]]);
                   delete(PENDING)
                 Else
                     Idle_Flag=True;
   }
               }
11. Else //TAM available < 0 or idle
   {
        Calculate Next_Schedule_Time = Finish[i], such that Finish[*i*]> This_Time and Finish[*i*] is minimum;
        Set This_Time=Next_Schedule_Time;
       12. For every Core i, such that finish[i] = This_Time
                 Wavail = Wavail + Width[*i*];
                 13. Set Complete[*i*] = TRUE;
        Idle_Flag=False;
   }
} //end of while

```
    return  test_schedule;
}
```
**Fig. 3  Proposed Test scheduling algorithm width TAM optimization**

Data structure test_schedule

    1. width[$i$]     //TAM width assigned to core $i$

    2. finish[$i$]     //end time of core $i$

    3. scheduled[$i$]  //boolean indicates core $i$ is scheduled

    4. start[$i$]     //begin time of core $i$

    5. complete[$i$]  //boolean indicates test for core $i$ has finished

    6. peak_tam[$i$]  //equals to MAX_TAM$_u$ of core $i$

**Fig. 4 Data structure for the test schedule**

Procedure update( i , w)

1. Let i be the core to be updated in the test schedule

2. Start[$i$]=Current_Time;

3. Set scheduled[$i$] = TRUE;

4. finish[$i$] = Current_Time + T$_i$(w);

5. width[$i$]=w;

6. Wavail=Wavail- w;

**Fig. 5 Data structure for the update algorithm**

Next, we describe our solution to *Problem-$_{RP}$2*.

**Data Structure**

    The data structure in which we store the TAM width and testing time values for the cores of the SOC is presented in Fig. 4. This data structure is updated with the begin times, end times, and assigned TAM widths for each core as the test schedule is developed.

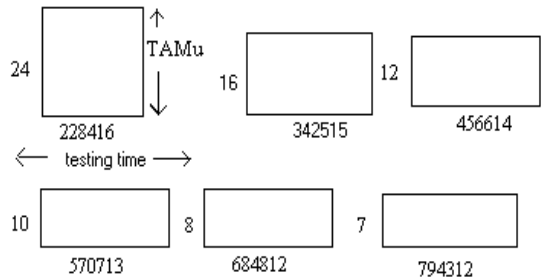

**Fig. 6 Example of some rectangles for core 6 of SOC p93791 (figure drawn not to scale) when W$_{max}$= 32**

**Rectangle Construction**

    In our proposed test scheduling algorithm (Fig. 3), after getting the result of Wrapper_Design, for each core, we construct a set of rectangles taking TAM$_u$ as rectangle height and its corresponding testing time as

rectangle width such that $TAM_u \leq W_{max}$ (Fig. 6) rather than constructing the collection of Pareto-optimal rectangles like[11]. $MAX\_TAM_u$ is the largest among the $TAM_u$ values satisfying the above constraint. In Fig. 7, $MAX\_TAM_u$=24 and $W_{max}$=32 .For combinational core, $MAX\_TAM_u$ is always equal to $W_{max}$. Note that, In case of TAM wire assignment to that particular scheduling of p93791 (Fig. 6), TAM wires that are to be assigned to core 6 must be selected from values 24,16,12,10,8-1 depending on TAM width available.

**Diagonal Length Calculation**

In line 2, we find the smallest ($T_{min}$) among the testing time corresponding to $MAX\_TAM_u$ for all cores. In line 3, for each core we divide width (testing time) of all constructed rectangles (line 3) with $T_{min}$. Then in line 4,for each core we calculate the diagonal length of the rectangle where rectangle height W[*i*] =$MAX\_TAM_u$ and rectangle width T[*i*] is reduced testing time corresponding to $MAX\_TAM_u$ .We sort the cores in descending order of diagonal length calculated in line 4 .

**TAM Assignment**

While executing the main **While** loop(line 7),if there are Wavail TAM wires available for assignment and list INITIAL is not empty, we select a core *c* from the list in sorted order. If TAM available at that moment, Wavail is greater than or equal to peak_tam[*c*] and there is no power conflict, we schedule the tests of that core and assign TAM wires to c equal to peak_tam[*c*].Note that ,peak_tam[*c*] is equal to $MAX\_TAM_u$ of core *c*. If Wavail is less than peak_tam[*c*] and power constraints is satisfied, it tries to find a $TAM_u$ value such that $TAM_u \leq$ Wavail and $TAM_u$ greater than half of peak_tam[*c*]. If it fails to assign TAM wires to c satisfying these conditions, it add the core *c* into queue PENDING.It then deletes a core p from the queue PENDING for scheduling only if Wavail is greater than or equal to peak_tam[*p*] and there is no power conflict.

If list INITIAL is empty, the algorithm deletes the core *c* at the front of queue PENDING only if Wavail $\geq$ peak_tam[*c*] and power constraints is satisfied. Otherwise it waits until sufficient TAM wires become available and power constraints are satisfied. If Wavail>0 and INITIAL is empty, these Wavail wires are declared idle and Idle_Flag is set if Wavail cannot satisfy power constraints as well as the condition Wavail $\geq$ peak_tam[*c*] where *c* is the core at the front of queue PENDING.

if there are Wavail idle wires or Wavail=0, the execution proceeds to line 12 where the process of updating This_Time to Next_Schedule_Time and Wavail is begun .Line 13 increases Wavail by the width of all cores ending at the new value of This_Time and Line 13 sets complete[*i*] to true for all cores whose test has completed at This_Time.

**EXPERIMENTAL RESULTS**

In this section, we present experimental results for one example SOC: d695. This SOC is a part of the ITC'02 SOC benchmarking initiative [4].In our algorithm we considered TAM wire sharing and power constraints as test conflict. Note that none of the previous approaches consider more test conflicts than TAM wire sharing but [6] which take power constraints, hierarchical constraints, precedence constraints, unit testing with multiple test sets into account.

In the ITC'02 benchmark specification, no power data are given for this system. Therefore, we add power values for each core depicted in Table 2. In the first experiment we compare our technique with the approach presented by [6] and [16] using the d695 circuit considering the same power values depicted in Table 2. The results are given in Table 3 and Table 4 for different TAM width.

In our second experiment, we compared our approach to previous proposed techniques without considering any power limitation. The results are for a range of TAM bandwidths given in Table 5.

| Core $C_i$ | $P_i$ |
|---|---|
| 1 | 660 mW |
| 2 | 602 mW |
| 3 | 823 mW |
| 4 | 275 mW |
| 5 | 690 mW |
| 6 | 354 mW |
| 7 | 530 mW |
| 8 | 753 mW |
| 9 | 641 mW |
| 10 | 1144 mW |

**Table 2. Power consumption values for d695**

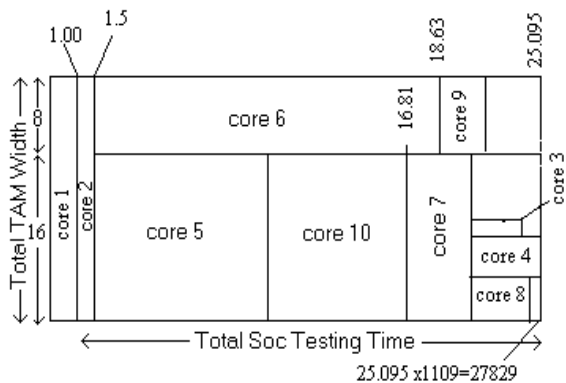

**Fig. 7** Test scheduling for d695 using our algorithm ($T_{min}$=1109 and TAM width=24) without power constraints

## CONCLUSION

In this paper we have proposed a Wrapper/TAM co-optimization and test scheduling technique that takes test power consumption into account when minimizing the test application time. It is important to consider test power consumption since exceeding it might damage the system. The proposed technique is based on rectangle packing which emphasizes on both time and TAM width by considering diagonal lengths. The experimental results show the efficiency of our algorithm.

| $P_{max}$ Approach | TAM Width=16 | | | TAM Width=24 | | | TAM Width=32 | | | TAM Width=40 | | |
|---|---|---|---|---|---|---|---|---|---|---|---|---|
| | [6] | [16] | Proposed | [6] | [16] | Proposed | [6] | [16] | Proposed | [6] | [16] | Proposed |
| 1500 | 45560 | 43541 | 40855 | 32663 | 31028 | 38705 | 26973 | 27573 | 21004 | 24369 | 20914 | 20856 |
| 1800 | 42450 | 44341 | 40855 | 32054 | 29919 | 33010 | 23864 | 24454 | 21004 | 18774 | 20467 | 22261 |
| 2000 | 42450 | 43221 | 39572 | 29106 | 29419 | 33010 | 21942 | 24171 | 21004 | 18691 | 19206 | 20978 |

**Table 3. Power constrained test time on design d695**

| $P_{max}$ Approach | TAM Width=48 | | | TAM Width=56 | | | TAM Width=64 | | |
|---|---|---|---|---|---|---|---|---|---|
| | [6] | [16] | Proposed | [6] | [16] | Proposed | [6] | [16] | Proposed |
| 1500 | 23425 | 20914 | 21473 | 19402 | 16841 | 18072 | 19402 | 16841 | 18163 |
| 1800 | 18774 | 18077 | 18966 | 18774 | 14974 | 16102 | 16804 | 14899 | 14041 |
| 2000 | 17467 | 17825 | 16868 | 14563 | 14128 | 16102 | 14469 | 14128 | 14914 |

**Table 4. Power constrained test time on design d695**

| TAM Width | [6] | [10] | [11] | [12] | [16] | Proposed |
|---|---|---|---|---|---|---|
| 64 | 13348 | 12941 | 11604 | 12941 | 11279 | 14914 |
| 48 | 17257 | 16975 | 15698 | 15300 | 15142 | 15075 |
| 40 | 18691 | 17901 | 18459 | 18448 | 17366 | 20254 |
| 32 | 20512 | 21566 | 23021 | 22268 | 21389 | 20402 |
| 24 | 29106 | 28292 | 30317 | 30032 | 28639 | 27829 |
| 16 | 41847 | 42568 | 43723 | 42644 | 42716 | 39572 |

**Table 5.Experimental result for d695**